\documentclass[prl,showpacs,twocolumn,superscriptaddress,floatfix]{revtex4}
\usepackage{bbm,hyperref,amssymb,amsmath,times,graphicx}
\newcommand{\ket}[1]{\vert #1 \rangle}

\newcommand{\tr}{\mathop{\mathrm{Tr}}\nolimits}
\newcommand{\var}{\textrm{Var}}
\newcommand{\E}{\mathbb{E}}

\newcommand{\Ord}[1]{O\left( #1 \right)}
\begin{document}
\title{Optimal quantum estimation of loss in bosonic channels}
\author{Alex Monras}
\email{amonras@ifae.es} \affiliation{Grup de Fisica Te\`orica \&
IFAE, Universitat Aut\`onoma de Barcelona, Bellaterra, Spain}
\author{Matteo G. A. Paris}
\email{matteo.paris@fisica.unimi.it}
\affiliation{Dipartimento di Fisica dell'Universit\`a di Milano, Italia.}
\date{\today}
\begin{abstract}
We address the estimation of the loss parameter of a bosonic channel probed
by Gaussian signals. We derive the ultimate quantum bound on precision and
show that no improvement may be obtained by having access to the
environment degrees of freedom. We found that, for small losses, the
variance of the optimal estimator is proportional to the loss parameter
itself, a result that represents a qualitative improvement over the shot
noise limit. An observable based on the symmetric logarithmic derivative is
derived, which attains the ultimate bound and may be implemented using
Gaussian operations and photon counting.
\end{abstract}
\maketitle
In all the branches of physics, many quantities of interest are not
directly accessible, either in principle like the measurement of all
fields \cite{Mab01} or due to experimental impediments.  In these
cases, one should resort to indirect measurements, inferring the
value of the quantity of interest from its influence on a given
probe. This is basically a parameter estimation problem whose
solution, {\em i.e} the determination of the most precise estimator,
unavoidably involves an optimization procedure \cite{Cra46}.
Indeed, the central result in classical estimation is the so-called
Cramer-Rao inequality, which sets a lower bound on the variance of
any estimator in terms of the Fisher information.
When quantum systems are involved, the optimal measurement to detect an
unobservable quantity may be found using tools from quantum estimation
theory \cite{Hel76,Hol82}. The quantum version of Cramer-Rao inequality has
been established \cite{Hel76,Hol82,Bra94,Bra96} and the lower bound imposed
by the quantum Fisher information has been shown to be achievable
asymptotically \cite{Bra96}. In turn, this results permit to write a parameter
based uncertainty relation for unobservable quantities \cite{Bra96,Mac06}.
\par
Quantum estimation theory has been mostly applied to find optimal
measurements and, in turn, to evaluate the corresponding lower bounds
on precision, for the estimation of parameters imposed by unitary
transformations. For bosonic systems these include single-mode phase
\cite{Hol79,Dar98,Mon06}, displacement \cite{Hel74}, squeezing
\cite{Mil94,Chi06} as well as two-mode transformations, e.g. bilinear
coupling \cite{Per01}.
Concerning open quantum systems and non unitary processes, QET has
been applied only to finite dimensional systems \cite{Hot06}, to
optimally estimate the noise parameter of depolarizing \cite{Fuj01}
or amplitude-damping \cite{Zhe06} channels. On the other hand, to
the best of our knowledge, QET has been not exploited to determine
the optimal way of estimating parameters of a lossy bosonic channel.
Needless to say, besides fundamental interest, quantum limits to the
estimation of bosonic lossy channels are extremely relevant for
applications, as for example absorption measurements and
characterization of optical media \cite{Dau06,Gra87,Pol92}.
\par
The precision of a measurement corresponds to the smallest value of
the parameter that can be discriminated. In several quantum
mechanical schemes, precision is improved by the more efficient use
of the resources employed in the measurement process
\cite{Dar01,Aci01}. A relevant example is the quadratic increase of
the precision in phase estimation by squeezed states. On the other
hand, despite the improvement in the resource-precision trade-off,
known estimation schemes are characterized by a threshold value,
under which the parameter cannot be measured. The results described
in this letter, besides being the first example of optimized
parameter estimations for non unitary maps, show a novel feature of
quantum-limited measurements: the asymptotic variance scales with
the parameter itself. Therefore, our results may be applied to
arbitrarily small values of the parameter, with a fixed relative
variance.
\par
The scheme we are going to consider is the following: the dynamics of a
bosonic quantum system is governed by a Lindblad Master equation of the form
\begin{align}
\dot{\varrho}=\frac{\gamma}{2}\mathcal L[a]\varrho\:,
\label{ME}
\end{align}
which results from the interaction of the system with an external environment,
as for example a bath of oscillators. We want to estimate the value of the
loss parameter $\gamma$, and to this purpose a set of identically prepared
signals (probe state) are sent through the channel and measured at the output.
We denote by $\varrho_0$ the input state and by $\varrho_\gamma = E_\gamma
(\varrho_0)$ the state at the output of the channel, $E_\gamma$ being the map
associated to the evolution (\ref{ME}). An estimation strategy for $\gamma$
consists of three ingredients: a suitable input signal, a measurement at the
output and an inference rule to extract the value of the parameter from the
experimental sample $\boldsymbol \chi$, {\em i.e.} the set of measurement
outcomes. The ultimate goal of choosing an estimation strategy is to find a
scheme that maximizes precision with the constraint of a finite fixed amount
of energy impinged into the channel. In particular, in this letter we address
the following questions: 1) Which is the best probe state? 2) Which is the
optimal measurement that should be performed at the output? 3) Which is the
attainable precision? 4) Can the precision be improved by accessing the
environment degrees of freedom, e.g. the bath of oscillators?
\par
The quantum state at the output of the channel may be formally written
as $\varrho_\gamma = E_\gamma (\varrho_0 ) = \exp\left(\gamma t/2 {\cal
L}[a]\right)\varrho_0$, whereas the constraint of finite probe energy reads
as follows $\hbox{Tr}[\varrho_0\, a^\dag a]=\bar n$, $[a,a^\dag]=1$ being
the bosonic mode operators. In this letter we focus attention on Gaussian
probe states and consider a zero-temperature environment without squeezing,
i.e. we focus on lossy channels described by a superoperator of the form
\begin{equation}
\mathcal L[a]\varrho=2a^\dagger \varrho a-a^\dagger a\varrho-\varrho a^\dagger a\:.
\end{equation}
Given the evolution time (length) t, the aim is to estimate as
precisely as possible the parameter $\gamma$ in the Lindblad Master
equation (\ref{ME}). In the following we consider the
parametrization $\exp(-\gamma t) = \cos^2\phi$ and seek for the best
estimation strategy for  $\phi$. We denote the probe state as
$\varrho_0$ and the evolved state as $\varrho_\phi$ with the Master
equation that rewrites as $d\varrho/d\phi = \tan\phi\, {\cal
L}[a]\,\varrho$.  Our goal is to determine the optimal probe state
$\varrho_0$ and the optimal measurement, i.e the probability
operator-value measure (POVM) $O_{\boldsymbol \chi}$ in order to
form the estimator $\hat\phi$ to infer the the value of $\phi$ from
the set of outcomes ${\boldsymbol \chi}$. Notice that, being our
estimation problem uniparametric, the optimal measurement can be
realized using individual measurements on separate (subsequent)
preparations \cite{Gil00}, so that the expectation value of the estimator
can be written as $\E_\phi[\hat\phi]=\sum_\chi \hat\phi({\boldsymbol
\chi})\tr[\varrho_\phi)^{\otimes N}\, O_{\boldsymbol\chi}]$.  We
focus attention on asymptotically unbiased estimators, \emph{i.e.}
those for which $\E_\phi[\hat\phi]\rightarrow\phi$ as
$N\rightarrow\infty$. Those are interesting because as the number of
experimental data increases, all systematic errors go to zero. The
variance of an unbiased estimator is bounded by the Cramer-Rao
inequality \cite{Cra46},
\begin{equation}
    \label{eq:Heisenberg}
    \var_\phi[\hat\phi]\geq\frac{1}{F(\phi)},
\end{equation}
where $F(\phi)$ is the Fisher information of the measurement.
$F(\phi)$ is additive, {\em i.e} the total Fisher information of
measurements on multiple copies of $\varrho_\phi$ is the sum of the
individual ones. It is worth mentioning that the so-called maximum
likelihood estimator asymptotically attains this bound for large
number of identical repeated samplings of a probability
distribution. In such case the asymptotic variance scales as $1/N$.
{\em i.e.} $\var_\phi[\hat\phi]\rightarrow [NF(\phi)]^{-1}$. The
Fisher Information $F(\phi)$ is bounded from above by the Quantum
Fisher Information (QFI) $F(\phi)\leq H(\phi)$ which, in turn,
provides a measure of the ultimate precision available with a given
quantum state. QFI is additive too and thus we have
\begin{equation}
    \var_\phi[\hat\phi]\geq \frac{1}{N H(\phi)}.
\end{equation}
where $H(\phi)$ can be expressed in terms of the
\emph{symmetric logarithmic derivative}, (SLD) $\Lambda(\phi)$.
SLD is implicitly defined as the Hermitian operator that satisfies
\begin{equation}
    \label{eq:derivative}
    \frac{d\varrho_\phi}{d\phi}=\frac{1}{2}\left[\varrho_\phi
    \Lambda(\phi)+\Lambda(\phi)\varrho_\phi\right],
\end{equation}
from which the QFI can be computed as
$H(\phi)=\tr[\varrho_\phi\,\Lambda(\phi)^2]$. $\Lambda (\phi)$ is an
observable which depends on $\phi$ and, analogously to the classical
LD, its expectation value is zero, \emph{i.e.}
$\tr[\Lambda(\phi)\, \varrho_\phi]=0$. A biased
$\Lambda(\hat\phi)$ reveals to which extent the measured state
differs from $\varrho_{\hat\phi}$.  In addition, it has been proved
\cite{Bra94} that an optimal measurement can be obtained by
projecting onto one-dimensional eigenspaces of $\Lambda(\phi)$,
\emph{i.e.} a measurement for which the Fisher information is
maximized, $F(\phi)=H(\phi)$. Upon this considerations, in the
following we consider the one-step adaptive strategy
\cite{Gil00,Mon06}, in which one makes a rough estimate $\hat\phi_0$
on a vanishing fraction of copies $N^{\delta}$, $1/2<\delta<1$, and
then measures $\Lambda(\hat\phi_0)$ on the remaining copies in order
to refine the estimate $\hat\phi$.
\par
Notice that Eq. \eqref{eq:Heisenberg} can be regarded as a
Heisenberg relation for parameter estimation. In the case of pure
states and unitary evolution of the form $U(\phi)=\exp (i G \phi)$
we have $H(\phi)=4\langle\Delta G\rangle^2$, where $\langle\Delta
G\rangle^2$ is the uncertainty of the generator $G$ on the state
used as a probe. Overall, we get the inequality
$\var_\phi[\hat\phi]\langle\Delta G\rangle^2\geq1/(4N)$, 
which represents a parameter-based uncertainty relation for unobservable
quantities \cite{Bra96,Mac06}.
\par
Any Gaussian state of a single bosonic mode may be represented as a
thermal state $\rho_\mu$ under the action of a squeezing and a
displacement operations {\em i.e.}
$\varrho=D(\alpha)S(\zeta)\rho_{\mu}
S^\dagger(\zeta)D^\dagger(\alpha)$, where $\alpha=s e^{i\theta}$ is
the displacement amplitude and $\zeta=r e^{-2i\varphi}$ is  the
squeezing parameter; $D(\alpha)=\exp(\alpha a^\dagger-\alpha^* a)$,
$S(\zeta )=\exp(\frac12 \zeta^2 a^{\dag 2} - \frac12 \zeta^{* 2}
a^2)$. The thermal state can be expressed in terms of its purity as
$\rho_\mu= 2\mu/(1+\mu) [(1-\mu)/(1+\mu)]^{a^\dagger a}$. At this
point, some considerations about the choice of a probe state are in
order. We aim at finding the Gaussian state $\varrho_0$ which, under
the action of the amplitude damping channel, is mapped onto
$\varrho_\phi$ with highest QFI. Of course, if an arbitrary amount
of energy were available, an infinite precision could be reached. If
instead one restricts to a probe with finite energy, then a
compromise between $|\alpha|$, $r$ and $\mu$ must be achieved. From
the phase symmetry of the amplitude damping channel, it is clear
that the only relevant angular parameter is the difference between
the displacement phase $\arg \alpha$ and the squeezing direction
$\arg\zeta$. Hence, we can take $\zeta\equiv r \in {\mathbbm R}$ and
consider $\theta=\arg \alpha$ as the relevant angle. On the other
hand, as also intuitively expected, it is of no use to spend energy
in preparing a thermal state, that is, for any given input
energy $\bar n$, the optimal probe state is pure.
Therefore, the optimization problem can be reduced to determine two
parameters, the the ratio $x$ of squeezing energy to total energy
and the displacement phase $\theta$. Since for a pure Gaussian state
$\varrho_0=D(\alpha_0)S(r_0)|0\rangle\langle 0| S^\dag (r_0)D^\dag
(\alpha_0)$ the mean photon number is given by $\bar
n=\sinh^2r_0+|\alpha_0|^2$ we have $\sinh^2r_0=x \bar n$ and
$|\alpha_0|^2=(1-x)\bar n$.
\par
The evolution of the state parameters
under the action of the channel may be explicitly evaluated
\cite{Ser05}
\begin{eqnarray}
\nonumber
\mu_\phi&=&1/\sqrt{\cos^4\phi+\sin^4\phi+2\cos^2\phi\sin^2\phi\cosh2r_0}\\
\nonumber
r_\phi&=&\tfrac{1}{2}\cosh^{-1}\left[\mu_\phi\left(\cos^2\phi
\cosh2r_0+\sin^2\phi\right)\right]\\
\label{evp}
s_\phi &=&s_0\cos\phi.
\end{eqnarray}
We can now evaluate the SLD $\Lambda (\phi)$, which corresponds to the optimal
measurement and allows to calculate the QFI. Upon writing $\varrho_\phi$ in its
diagonal form, $\varrho_\phi = \sum_k \varrho_k |\psi_k\rangle\langle \psi_k
|$, $|\psi_k\rangle = D(\alpha) S(r) |k\rangle$ one easily finds
\begin{align}
\Lambda (\phi) = 2 \tan\phi\, \sum_{pq} \frac{\langle \psi_q | {\cal
L}[a]\varrho|\psi_p\rangle}{\varrho_p + \varrho_q}\, |\psi_q\rangle\langle
\psi_p|\:.
\end{align}
A lengthy but straightforward calculation yields
\begin{align}
\Lambda (\phi) & = 2 \tan\phi D(\alpha) S(r) K S^\dag (r) D^\dag (\alpha) \\
K &= \left[ A\, a^\dag a + B (a^2 + a^{\dag 2}) - C a -C^* a^\dag + F\right]
\nonumber
\end{align}
where
\begin{align}
A & = \frac{2\mu}{1-\mu^2}(\mu\cosh2r-1)
\quad B = \frac{\mu^2}{1+\mu^2}\sinh 2r \nonumber \\
C & = \mu (\alpha \cosh r + \alpha^* \sinh r) \quad F = 1- \frac{2\mu\cosh^2
r}{1+\mu^2}\:.
\label{pars}
\end{align}
All the parameters in (\ref{pars}) are given in Eq. (\ref{evp}), {\em i.e.}
refer to the state $\varrho_\phi$ after the evolution in the  lossy
channel. We omitted the explicit dependence on $\phi$ for brevity.  Using
the identity
\begin{align}
K = F' + \tau S^\dag (\eta) D(\beta)\, a^\dag a\, D(\beta) S(\eta)
\end{align}
where $F' = F + \frac12 (\tau - A - 2 \tau |\beta|^2)$, $\tan 2\eta
= - 2B/A$, $\tau=\sqrt{A^2 - 4 B^2}$ and $\beta = (C \cosh r + C^*
\sinh r)/\mu\tau$, one sees that the eigenvectors of $\Lambda
(\phi)$ are of the form
$D(\alpha)S(r)S^\dagger(\eta)D^\dagger(\beta)\ket{n}$, which means
that the measurement of $\Lambda (\phi)$ may be implemented with
Gaussian operations and photon counting, e.g. by squeezing and
displacing the state under investigation and then measuring the
photon number distribution by a suitable reconstruction technique
\cite{Zam05}.
\par
\begin{figure}[h]
\includegraphics[width=0.22\textwidth]{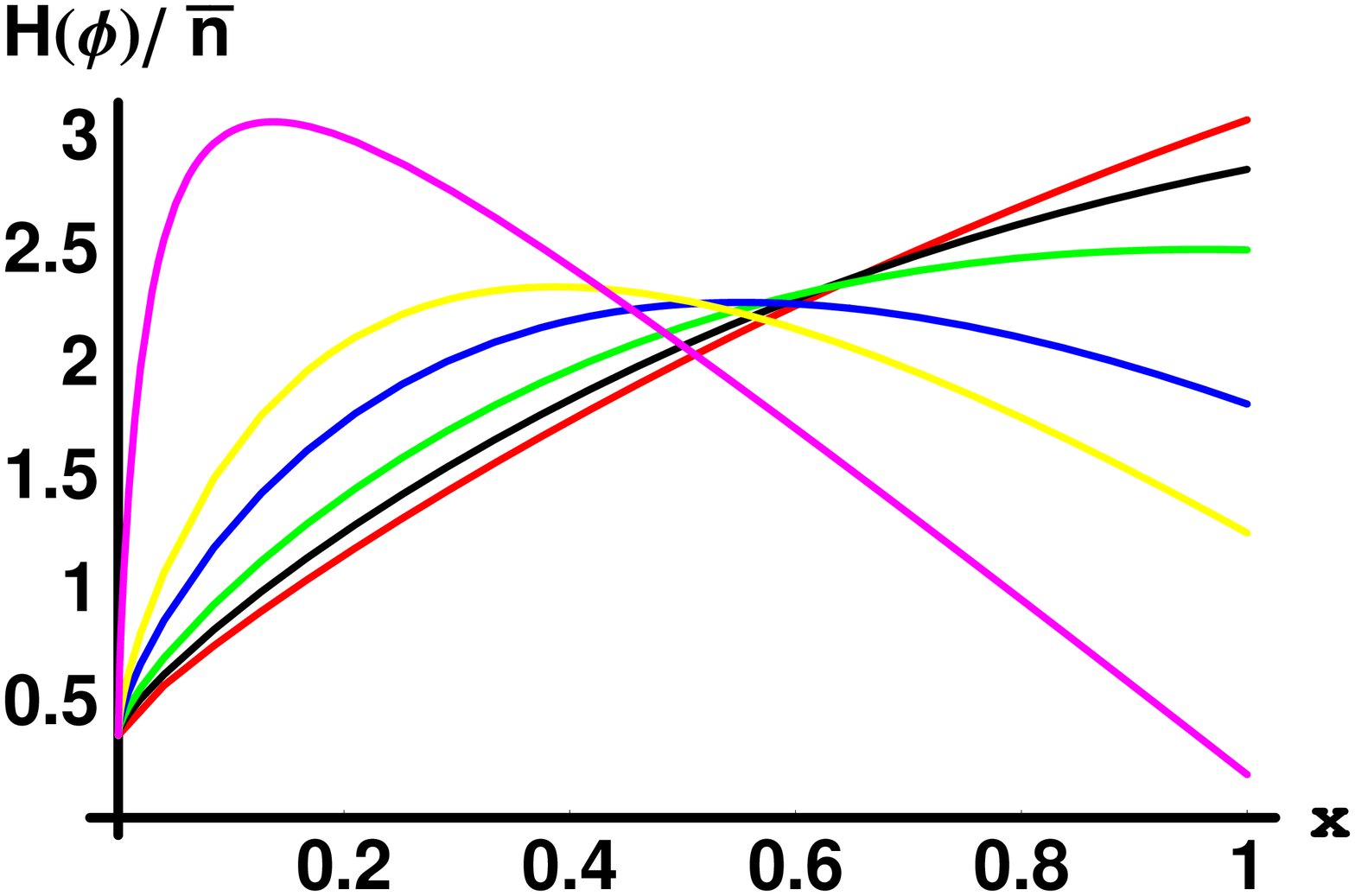}
\includegraphics[width=0.22\textwidth]{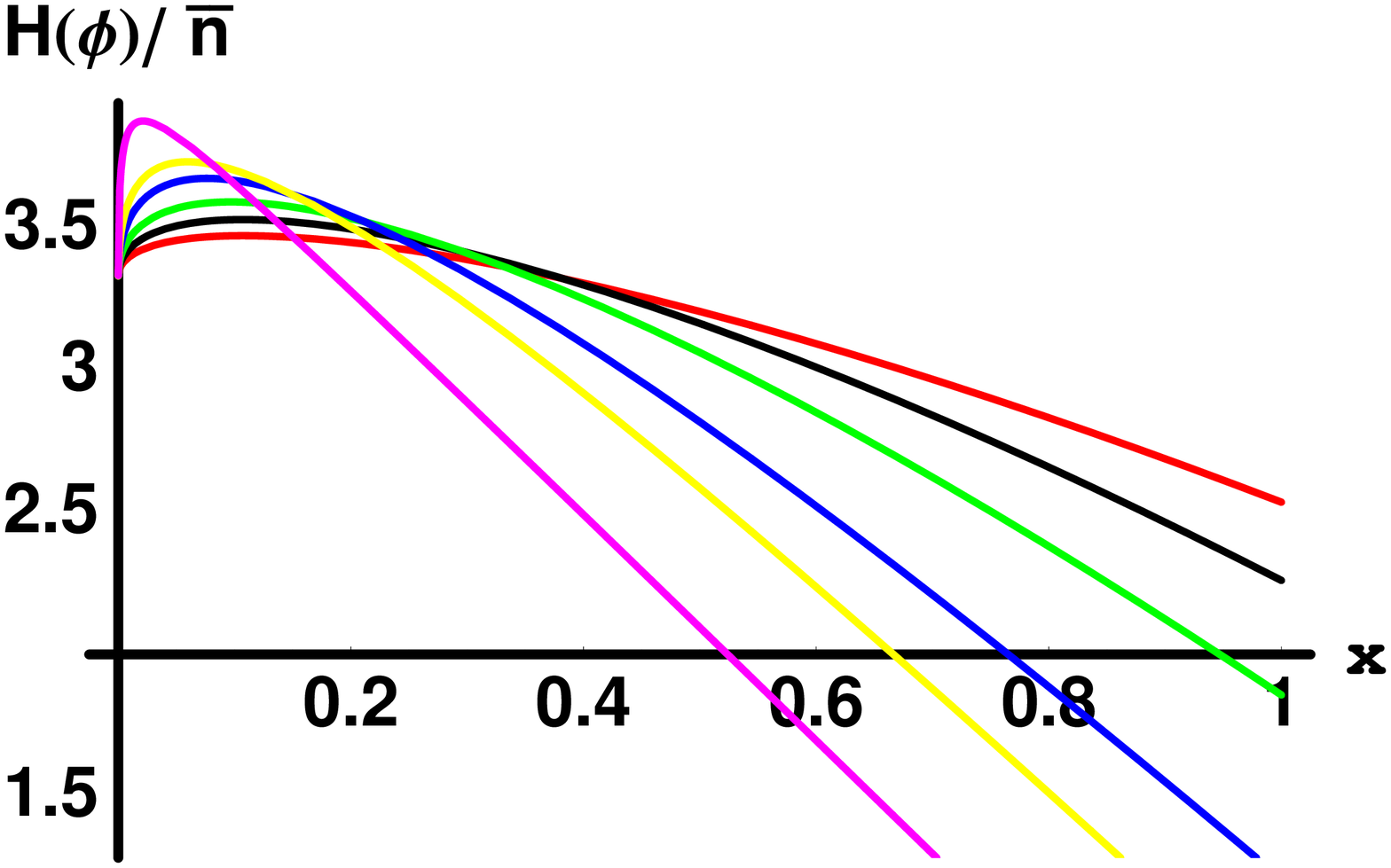}
\caption{Normalized Quantum Fisher information $H(\phi)/\bar n$ as a function of the
squeezing fraction of the probe state for different values of the
probe energy and two values of the actual loss parameter. (Left):
$\tan^2\phi=0.1$. (Right): $\tan^2\phi=5$. In both plots, from bottom to
top in the region $x\simeq 0$, the curves for $\bar n=0.5 $, $\bar n=1$,
$\bar n=2$, $\bar n=5$, $\bar n=10$, and $\bar n=100$.\label{f:HR}}
\end{figure}
\begin{figure}[h]
\includegraphics[width=0.23\textwidth]{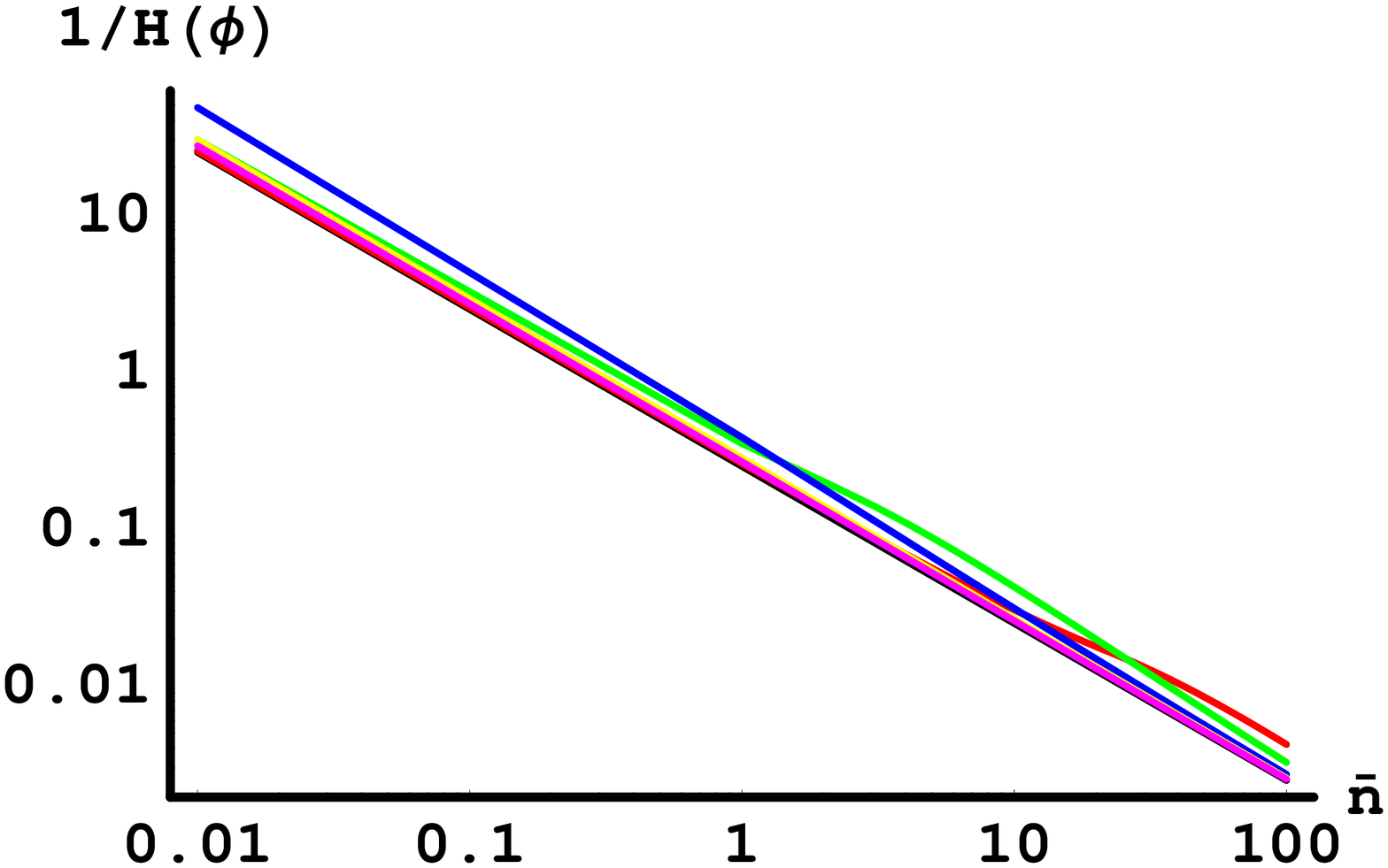}
\includegraphics[width=0.22\textwidth]{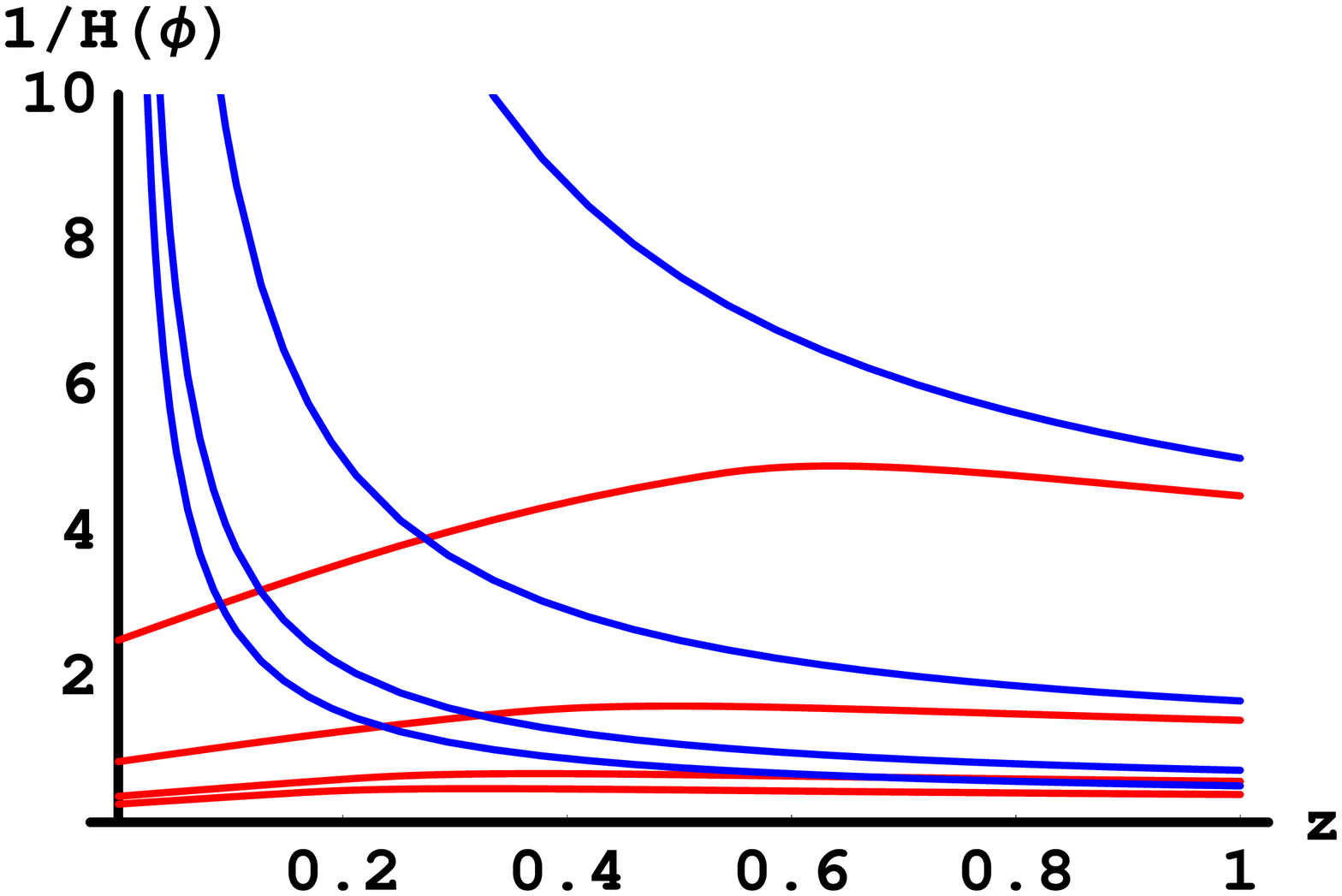}
\caption{(Left): Rescaled optimal variance $H^{-1}(\phi)$ as a
function of the probe energy for different actual values of the loss
parameter:  $z=0.01, 0.1, 1, 5, 10$. The larger is
$z$ ($\phi$) the closer is the curve to the asymptotic $H^{-1}(\phi )
= (4 \bar n)^{-1}$. (Right):
Rescaled optimal variance compared to the variance obtained for coherent
probe as a function of the loss parameter for different values of the
probe energy: from top to bottom the curves for $\bar n=0.1, 0.3, 0.7, 1$
respectively. The improvement obtained using optimized squeezed states is
apparent.  \label{f:VAR}}
\end{figure}

The explicit evaluation of the QFI $H(\phi)$ yields
\begin{align}
H(\phi)&= \frac{4 z \bar n}{1+ z (2 + z + 4 \bar n x)}
\left[1 - x + 2 \bar n x + \frac{x}{z} + z \right.\\
& - \left. \frac{4 \bar n x^2 z (1+\bar n x)}{1+z (2+ z + 2 \bar n x)}
+ 2 (1-x)\sqrt{\bar n x (1 + \bar nx)}\right]\:, \nonumber
\end{align}
where $z = e^{\gamma t}-1 = \tan^2\phi$, and where we have already
performed the trivial optimization over the phase $\theta$, which yields
$\theta=0$, i.e. the displacement should be performed along the same
direction of squeezing. The optimization procedure thus reduces to
maximizing the QFI $H(\phi)$ with respect to the squeezing ratio $x$. In
Fig. (\ref{f:HR}) we report the renormalized QFI $H(\phi)/\bar n$ as a
function of the squeezing ratio for different values of $\bar n$ and of the
actual loss parameter. As it is apparent from the plots in the regime of
small losses and small probe energies the optimal probe is the squeezed
vacuum ($x_{opt} =1$), whereas for increasing energy there is a nonzero
value of the optimal squeezing fraction $x_{opt}$, which is a monotonically
decreasing function of both the probe energy and the loss parameter itself.
In the small energy regime, with squeezed vacuum as the optimal probe, the
QFI reads
\begin{align}
H(\phi) = \frac{4 \bar n (1+z^2)}{1+2 z(1+\bar n) + z^2} \simeq 4 \bar n +
\Ord{\phi^2}\:,
\end{align}
where the second equality expresses the attainable precision in estimating
the loss parameter for weakly damping channels. In the following, we will
see that this is the ultimate limit even if one has access to the
environment degree of freedom. In terms of $\gamma$ the bound  reads
\begin{equation}
    \var_\gamma[\hat\gamma]\rightarrow\frac{\gamma}{\bar n N t}+\Ord{\gamma^2}.
\end{equation}
This is a remarkable result since it is valid for any value of the
loss parameter with no lower bound (see also Fig. \ref{f:VAR}).
Recall, however, that squeezed vacuum probes will not be optimal for
large enough values of $\gamma$.
\par
The maximization of the QFI in the general case may be
done numerically. In Fig. \ref{f:VAR} (left) we report the log-log plot of
the rescaled optimal variance $ N \hbox{Var}_\phi [\hat \phi]= 1/H_{max}
(\phi)$ as a function of the probe energy $\bar n$ for different values
of the actual loss parameter. As it is apparent from the plot the variance
does not dramatically depend on the actual value of the loss parameter.
The common scaling is given by $\hbox{Var}_\phi [\hat \phi] \propto (\bar n
N)^{-1}$.
\par
In order to better understand the behavior in the large energy regime an
asymptotic analysis is in order. Upon expanding the optimization equation
$dH/dx=0$ around $z=0$ and retaining the leading order $z^{-1}$ we
obtain analytically the optimal value of $x$ for small $\gamma$, which
for large $\bar n$ can expressed as $x_{opt} = (4 \bar n z)^{-1/2}+ \Ord{1/n}$.
This shows that, when a large amount of energy is available,
one can improve performances, besides squeezing, by employing a fraction of
energy to displace the probe. The QFI in this regime reads
$H(\phi)=4z^{-1}\left(1-\sqrt{nz}\right)+(2+4n)+2z$,
which corresponds to a variance
$\var_\phi[\hat\phi]\rightarrow\frac{\phi^2}{4N}
\left(1+\sqrt{n}\phi\right)+\Ord{\phi^4}$
Finally, by rewriting this in terms of $\gamma$ we have
\begin{equation}
\var_\gamma[\hat \gamma]
\rightarrow\frac{\gamma}{\bar n N t}+\Ord{\sqrt\frac{\gamma}{(\bar nt)^3}},
\end{equation}
that is, the proportionality of the variance to the parameter is recovered
for large $\bar n$. Notice that the scaling $H(\phi)\propto \bar n$ is
valid also in the general case, as it is apparent from Fig. \ref{f:VAR}
(left).
\par
It is worth stressing the improvement of precision with respect to
that attainable with coherent states (shot-noise limit). In fact, by
taking $x =0$ one easily sees that in this case the QFI reads
$H(\phi)=4\bar n z/(1+z)$ which corresponds to to $\var_\gamma[\hat
\gamma]\rightarrow (\bar n N t^2)^{-1}\:$. In other words, the
proportionality of the variance to $\gamma$ cannot be achieved using
coherent probe states. In Fig \ref{f:VAR} (right) we report the
optimal rescaled variance $H^{-1}(\phi)$ as a function of the loss
parameter compared to the variance that can be obtained using
coherent probes. The improvement at small values of the loss
parameter is apparent.
This can be intuitively understood as
follows: for small values of $\gamma$ the action of the loss map
$E_\gamma$ on a coherent probe is only that of a small displacement,
whereas a squeezed vacuum is dramatically ``mixed up''. Hence, a
squeezed state is more sensitive to small losses. Contrarily, for
large values of $\gamma$ a coherent state is highly displaced
whereas a squeezed vacuum state becomes close to pure again, losing
its sensitivity to $\gamma$.
\par
Let us now discuss other ways in which this performance could, in
principle, be improved. The Heisenberg limit generally describes the
ultimate precision attainable for parameter estimation. As we have seen
above, this holds in a strict sense when one deals with unitary
transformations, where the QFI characterizes the sensitivity of a state for
the estimation of a parameter. However, when one deals with non-unitary
transformations, this may not be true because the probe state evolves into
an entangled state with the environment. The access to the degree of
freedom of the environment may provide improved precision to the
measurement. In the case of the loss of a channel, some of the energy
present in the probe state is lost through coupling to the environment. The
Master Equation (1) can be seen as the effective interaction of the mode
$a$ with a second vacuum mode $b$ through a bilinear (beam-splitter-like)
evolution of the form $U(\phi)=\exp\left[i \phi\left(a^\dagger b+
a b^\dagger\right)\right]$.
This allows to have a unitary representation of the process. Here we use
this picture to derive the attainable precision in the hypothetic case that
one had access to the bath of oscillators.  In such a situation, the state
under inspection would remain pure, therefore $\Lambda(\phi)=2d\rho/d\phi$,
as can be seen by taking the derivative of the identity $\rho^2=\rho$. 
The generator $G$ is given by
$G=(a^\dagger b+a b^\dagger)$, and the uncertainty $\langle\Delta
G\rangle^2$ is $\langle\Delta G\rangle^2=\bar n$. As a consequence one gets
$\var_\phi[\hat\phi]\rightarrow\frac{1}{4\bar n N}$, which corresponds to
the precision attained using squeezed vacuum probe.
\par
In conclusion, we have shown that using Gaussian squeezed probes one can
improve the estimation of the loss parameter of a bosonic noisy channel. 
As it holds for any single parameter
quantum estimation problems, a one- step adaptive scheme is optimal to
leading order, thus providing a practical way for implementation.  We have
shown that the optimal measurement for Gaussian probe states can be
implemented by means of Gaussian operations and photon counting.
Furthermore, for small losses, the estimator variance obtained by choosing
appropriate probe states decreases proportionally to the loss parameter,
thus providing unlimited resolution for arbitrary small losses. We have
also obtained the optimal trade-off between squeezing and displacement of
probe states, showing that squeezed vacuum states are optimal in the small
energy limit. Finally, we have shown that even by having access to the
environment, one cannot improve the performance of our scheme.
\par\noindent\\
A.M. would like to acknowledge useful discussions with E. Bagan.
This work has been supported by MIUR project PRIN2005024254-002.
MGAP is also with ISI Foundation, Torino, Italy.

\end{document}